\documentclass[aps,prl,preprint,amsmath,amssymb,graphicx,longbibliography]{revtex4-1}

\usepackage{amssymb} 
\usepackage{amsmath}  
\usepackage{amsthm}
\usepackage{lgreek}
\usepackage{bm}
\usepackage{color}
\usepackage[usenames,dvipsnames]{xcolor}
\usepackage{graphicx}

\usepackage[squaren]{SIunits}

\begin{document}

\title{The topography of the environment alters the optimal search strategy for active particles}

\author{Giorgio Volpe}
\affiliation{Department of Chemistry, University College London, 20 Gordon Street, London WC1H 0AJ, United Kingdom}
\affiliation{Corresponding author: g.volpe@ucl.ac.uk}

\author{Giovanni Volpe}
\affiliation{Department of Physics, University of Gothenburg, 41296 Gothenburg, Sweden}

\date{\today{}}

\begin{abstract}
In environments with scarce resources, adopting the right search strategy can make the difference between succeeding and failing, even between life and death. At different scales, this applies to molecular encounters in the cell cytoplasm, to animals looking for food or mates in natural landscapes, to rescuers during search-and-rescue operations in disaster zones, and to genetic computer algorithms exploring parameter spaces. When looking for sparse targets in a homogeneous environment, a combination of ballistic and diffusive steps is considered optimal; in particular, more ballistic L\'evy flights with exponent $\alpha \le 1$ are generally believed to optimize the search process. However, most search spaces present complex topographies. What is the best search strategy in these more realistic scenarios? Here we show that the topography of the environment significantly alters the optimal search strategy towards less ballistic and more Brownian strategies. We consider an active particle performing a blind cruise search for non-regenerating sparse targets in a two-dimensional space with steps drawn from a L\'evy distribution with exponent varying from $\alpha =1$ to $\alpha = 2$ (Brownian). We demonstrate that, when boundaries, barriers and obstacles are present,  the optimal search strategy depends on the topography of the environment with $\alpha$ assuming intermediate values in the whole range under consideration. We interpret these findings using simple scaling arguments and discuss their robustness to varying searcher's size. Our results are relevant for search problems at different length scales, from animal and human foraging, to microswimmers' taxis, to biochemical rates of reaction.
\end{abstract}

\maketitle

\section*{Introduction}

What is the best strategy to search for randomly located resources? This is a crucial question for fields as diverse as biology, genetics, ecology, anthropology, soft matter, computer sciences and robotics \cite{BenichouRMP2011, ViswanathanBook2011}. In order to describe and analyse how a searcher browses the search space, many different plausible models have been proposed, including Brownian motion, intermittent search patterns, as well as L\'evy flights and walks \cite{BenichouRMP2011, ViswanathanBook2011, Shlesinger1986}. In particular, L\'evy statistics, among others models \cite{BenichouRMP2011}, have been successfully used to describe the emergence of optimal search strategies in natural systems at different length scales, from molecular entities \cite{LomholtPRL2005,ChenNatMat2015}, to swimming and swarming microorganisms \cite{ArielNatCom2015, KorobkovaNature2004, TuPRL2005}, to crawling eukaryotic cells \cite{HarrisNature2012}, to different species of foraging animals \cite{ViswanathanNature1996, ViswanathanNature1999, AtkinsonOikos2002, Ramos-FernandezBES2004, SimsNature2008, HumphriesNature2010, JagerScience2011}, to human motion patterns \cite{BrockmannNature2006, GonzalezNature2008, RaichlenPNAS2014}, although in the field of movement ecology there is some controversy on how universal L\'evy searches are \cite{EdwardsNature2007, MashanovaJRSI2010, Petrovskii2PNAS2011, HumphriesPNAS2012, JansenScience2012, ReynoldsPLR2015}. L\'evy statistics have also found applications in science and engineering, e.g., for  defining the optimal search strategy for robots \cite{DartelCS2004} and for describing anomalous diffusion and navigation on networks \cite{RiascosPRE2012,GuoSciRep2016}. 

The strategies based on L\'evy statistics can be described under a unified framework where the searcher is an active particle \cite{BechingerRMP2016} that performs random jumps (blind search) whose lengths $\ell$ are drawn from a stable distribution $P(\ell)$. The two limiting cases for $\alpha \to 0$ and $\alpha = 2$ correspond to ballistic and Brownian motion, respectively. The intermediate cases combine diffusive (i.e. local exploration) and ballistic (i.e. decorrelating, long-range excursions) steps in different proportions. In particular, the case for $\alpha = 1$ corresponds to a compromise superdiffusive regime, where the searcher explores its surroundings while reducing oversampling compared with a pure Brownian strategy \cite{ViswanathanNature1999, LomholtPNAS2008, ViswanathanBook2011}. When resources are plentiful, the most efficient strategy is a Brownian search ($\alpha = 2$)\cite{ViswanathanNature1999,SimsNature2008,HumphriesNature2010}; when resources are sparse, however, a L\'evy strategy with $\alpha = 1$ performs better over a pure Brownian strategy \cite{ViswanathanBook2011}. In general, more ballistic search strategies (i.e. $\alpha \le 1$) have been shown to be optimal in a wide range of situations, with the specific value of the exponent $\alpha$ dependent on, e.g., the nature of the encounters with the targets (i.e. destructive or non-destructive) or the presence of memory in the searcher's motion \cite{ViswanathanNature1999, BartumeusPRL2002, RaposoPRL2003, JamesPRE2008, RaposoPLOS2011, HumphriesPNAS2012, FerreiraPhysicaA2012}. 

These studies have limited their analysis to landscapes characterized by a barrier-free homogenous topography. However, in realistic scenarios the environment is often characterized by a more complex topography, where boundaries, barriers and obstacles play a crucial role in determining the searcher's motion. Examples of complex search spaces include cytoplasm for molecules within cells \cite{BarkayPhysToday2012}, biological tissue (or soil) for motile bacteria \cite{KimNatMicrobiol2016}, and patchy landscapes for foraging animals \cite{BoyerPRSLB2006}. This complexity can significantly influence the long-term behaviour of the system under study \cite{PinceNatCom2016}. As it has been recently shown, even a small perturbation, such as an external drift, can shift the optimal search strategy towards more Brownian strategies \cite{PalyulinPNAS2014}. 

Here, by considering a searcher performing a blind cruise search for uniformly distributed non-regenerating sparse targets, we show with numerical simulations and simple scaling arguments that the exponent that optimizes the search strategy depends on the topography of the environment. In particular, we show that, differently from the homogeneous case where typically $\alpha \le 1$ optimizes the search process, the optimal search strategy tends towards less ballistic and more Brownian cases, corresponding to values for the exponent $\alpha$ in the range  $(1,2]$.

\section*{Results}

\subsection*{Search in a homogenous topography}

\begin{figure}[!tbhp]
\centering
\includegraphics[width=11cm]{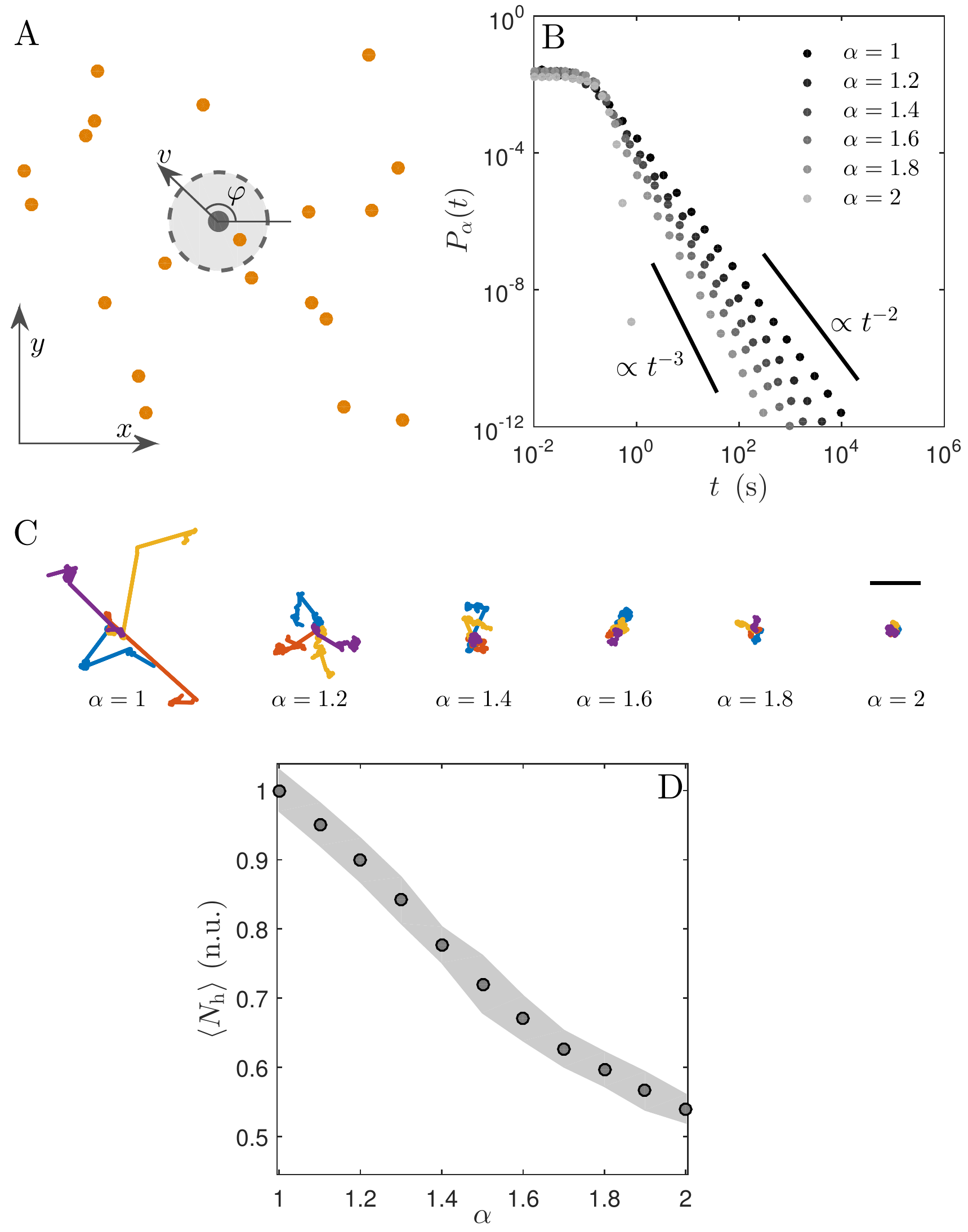}
\caption{\textbf{Optimal search strategy in a homogenous topography}. (A) Schematic representation (not to scale) of an active particle of radius $R$ blindly searching for uniformly distributed targets (dots) in a homogenous environment. The particle placed at position $[x,\, y]$ moves with constant speed $v$ and variable orientation $\varphi$. The capture radius is $r_{\rm c}$ (grey shaded area). (B) The time intervals $t_n$ with $n=0,1,2,...$ between changes of orientation $\varphi$ are drawn from a L\'evy distribution $P_{\alpha}(t)$ of exponent $\alpha \in [1, \, 2]$. The solid lines represent power-laws of exponent  $-\mu = -(\alpha+1)$ for the two limiting cases at $\mu = 2$ ($\alpha = 1$) and $\mu = 3$ ($\alpha = 2$). Note that for the case $\alpha = 2$ the distribution is a Gaussian, which is not a power law asymptotically. (C) Four different $1000$-s trajectories with a common origin are shown for various values of $\alpha$. The black scale bar corresponds to $1000 \, R$. (D) Average number of caught targets (circles) as a function of $\alpha$ in normalized units (n.u.). The values are averaged over $1000$ $1$-hour trajectories and normalized to the maximum value at $\alpha = 1$. The grey shaded area represents one standard deviation around the average values.}
\label{fig1}
\end{figure}

We start by analyzing an active particle of radius $R$ blindly searching for targets in an environment with a homogeneous topography, i.e. without any physical obstacles. As the active particle cruises the search space, it continuously captures the targets that come within a capture radius $r_{\rm c} = 2R$ from its center, as schematically shown in Fig.~\ref{fig1}A. 
The number of targets caught in each run is proportional to the area swept by the capture region surrounding the active particle. We assume the targets to be uniformly distributed, non-regenerating and scarce, i.e. with density $\rho \ll r_{\rm c}^{-2}$. The latter condition implies that, once an active particle captures a target, the probability of finding a second one is negligible if the particle moves by $\ell \lesssim r_{\rm c}$. 

The active particle performs a run-and-tumble motion, i.e. it has a fixed speed $v$ and changes its orientation $\varphi$ by a normally-distributed angle with zero mean and standard deviation $\sigma_\varphi$ at discrete time intervals $t_n$ with $n=0,1,2,...$ \cite{BechingerRMP2016}. In the following, we set $v = 5 R\,\, {\rm s}^{-1}$ and $\sigma_{\varphi} = \frac{\pi}{6}$. The time intervals $t_n$ between changes of direction are drawn from a L\'evy distribution $P_{\alpha}(t)$ of exponent $\alpha \in [1, \, 2]$ (Fig.~\ref{fig1}B) \cite{ViswanathanBook2011}. Asymptotically, this distribution tends to a power law with exponent $-(\alpha+1)$ for $\alpha \in [1,2)$ \cite{ViswanathanBook2011}:
\begin{equation}\label{e1}
P_{\alpha}(t) \approx A(\alpha) t^{-(\alpha+1)} \mbox{ for } t \rightarrow \infty,
\end{equation}
where $A(\alpha)$ is a normalization constant such that $\int_{0}^{\infty} P_{\alpha}(t) dt = 1$; for $\alpha = 2$ the distribution is a Gaussian which decays exponentially in $t$. As $v$ is constant, the run lengths $\ell_h(t_n) = v t_n$ are also distributed according to a L\'evy distribution of same index $\alpha$, thus leading the particle to move superdiffusively for $\alpha < 2$ and diffusively for $\alpha = 2$ at long times (Fig. S1A)\cite{ViswanathanBook2011}. Examples of trajectories for various values of $\alpha$ are shown in Fig.~\ref{fig1}C: as $\alpha$ decreases from the case of a pure Brownian strategy ($\alpha = 2$), the searchers tend to move ballistically over longer distances before a change in orientation occurs. These different superdiffusive regimes allow the searcher to explore the overall search space combining ballistic and diffusive steps in different proportions  \cite{ViswanathanNature1999,LomholtPNAS2008,ViswanathanBook2011}. Figure~\ref{fig1}D plots the average number of caught targets $\langle  N_{\rm h}\rangle$ obtained from 1000 simulated 1-hour trajectories as a function of $\alpha$. This number decreases as a function of $\alpha$, so that the optimal search strategy is for $\alpha = 1$ while the worst is the Brownian ($\alpha = 2$), in agreement with foraging theory \cite{ViswanathanBook2011}. 

\subsection*{Search in a porous topography}

\begin{figure}[!htbp]
\centering
\includegraphics[width=16cm]{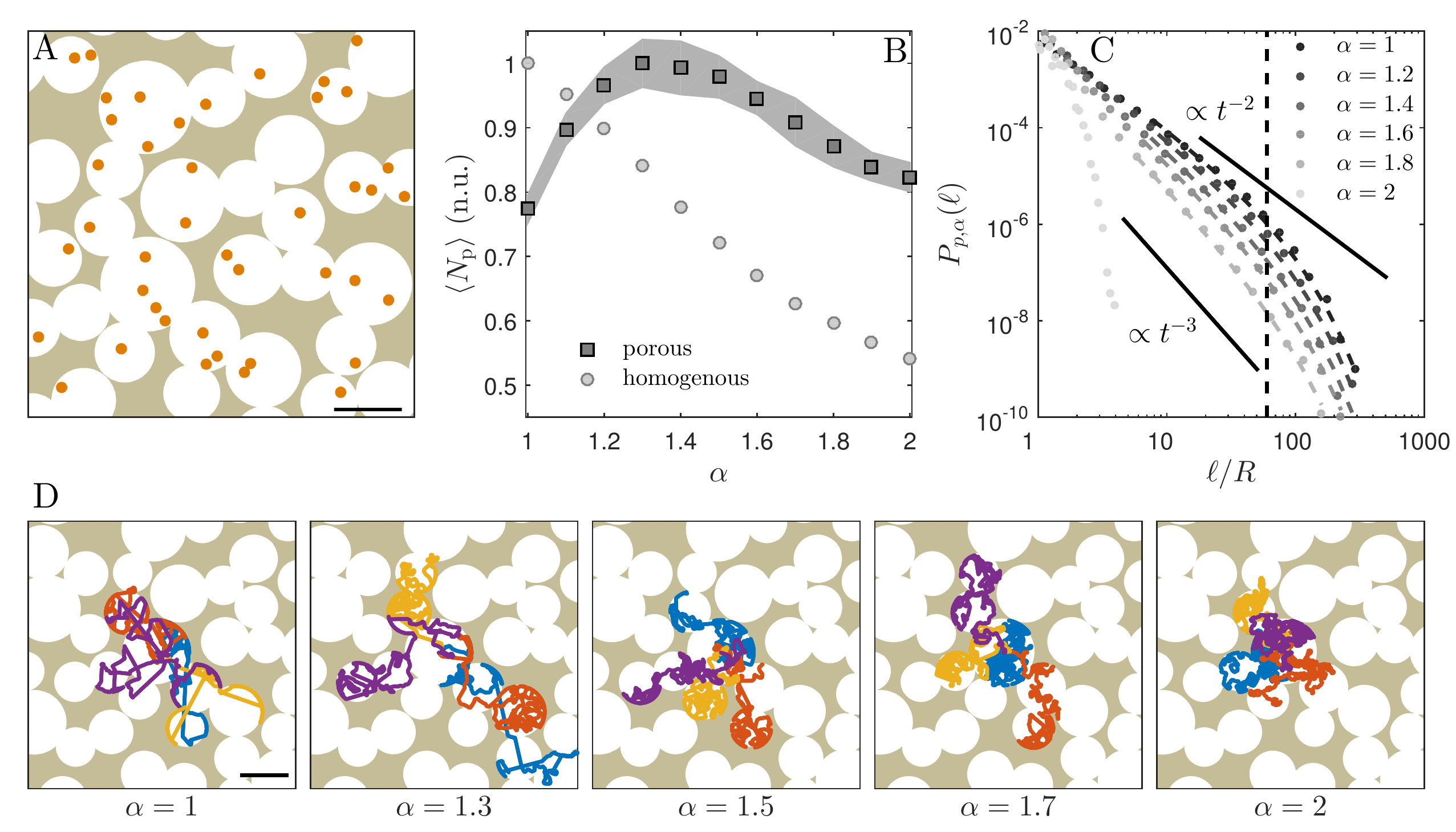}
\caption{\textbf{Shift of the optimal search strategy in a porous topography}. (A) A sample area of an extended two-dimensional porous medium where an active particle searches for uniformly distributed targets (dots). The porous medium is constituted by circular interconnected pores of average radius $R_{\rm p}$ ($R_{\rm p}/r_{\rm c} \approx 12.5$). (B) Average number of caught targets $\langle N_{\rm p} \rangle$ (squares) as a function of $\alpha$ in normalized units (n.u.). The values are averaged over $1000$ $1$-hour trajectories and normalized to the maximum value at $\alpha = 1.3$. The grey shaded area represents one standard deviation around the average values. To directly compare with the homogenous case, the trend of Fig.~\ref{fig1}D is also shown (circles). (C) Simulated probability distribution  of the run lengths $P_{{\rm \rm p},\alpha}(\ell)$ in the porous medium as a function of $\alpha$ (dots). The distributions are fitted to a power law with an exponential cutoff for $\alpha \in [1,2)$ (Eq.~\ref{e3}, dashed lines). The vertical dashed line represents the cutoff $R_{\rm c}$. (D) Four different $1000$-s trajectories with a common origin are shown for different values of $\alpha$. All black scale bars correspond to $50 \, R$.}
\label{fig2}
\end{figure}

In order to understand how the complexity of the environment influences the optimal search strategy, we now consider an active particle looking for targets in a medium with a heterogenous topography (Fig.~\ref{fig2}A). Specifically, the search space is now a two-dimensional porous medium constituted by uniformly distributed circular interconnected pores with average radius $R_{\rm p} \gg r_{\rm c}$; the characteristic size of a cluster of pores is much bigger than the total particle's displacement within the simulation time. We model the interaction with the pore walls using reflective boundary conditions so that the particle moves along the walls until its orientation changes to point away from the boundary \cite{VolpeAJP2014}. This scenario is realistic at different length scales as indeed both biological and artificial microswimmers and elementary robots behave in a similar way \cite{BechingerRMP2016,DimidovANTS2016}.

As it can be seen in Fig.~\ref{fig2}B, moving in such a porous environment shifts the optimal search strategy towards a more Brownian strategy ($\alpha = 1.3$) from the more ballistic case in the homogenous topography ($\alpha = 1$). This shift can be understood in quantitative terms by looking at the effective probability distribution of the run lengths $P_{{\rm \rm p},\alpha}(\ell)$ in the porous medium (Fig.~\ref{fig2}C). This distribution is well-approximated by a power law with an exponential cutoff at $R_{\rm c} = \lambda R_{\rm p}$
\begin{equation}\label{e3}
P_{{\rm \rm p},\alpha}(\ell) \approx B(\alpha,v) \ell^{-(\alpha+1)} e^{-\frac{\ell}{R_{\rm c}}} \mbox{ for } \ell \rightarrow \infty
\end{equation}
where $B(\alpha,v)$ is a normalization constant such that $\int_{0}^{\infty} P_{{\rm \rm p},\alpha}(\ell) d\ell = 1$ and $\lambda$ is a proportionality constant; $\lambda \approx 2.4$ is estimated by fitting the previous function to the simulated data and, in general, depends on the geometrical features of the medium. As a result of this interaction with the boundaries, therefore, the porosity affects longer run lengths more than shorter ones, thus mainly penalizing the more ballistic strategies over the more Brownian ones. In other terms, even if the changes in the particle's orientation are still dictated by the distributions in Fig.~\ref{fig1}B, the boundaries effectively limit the maximum run length leading the particles to perform a subdiffusive motion rather than a superdiffusive one as in the homogenous environment (Fig. S1B); this behavior is in accordance with observations on persistent random walkers in the presence of obstacles \cite{ChepizhkoPRL2013sub}. Qualitatively, this can also be appreciated by looking at some sample trajectories for different values of $\alpha$ (Fig.~\ref{fig2}D): when $\alpha$ decreases, the particles tend to spend longer portions of their trajectories at the walls, thus exploring less efficiently the inner area of the pores. It is interesting to note that, at least when the searcher explores the complex topography for a finite time as in our simulations, the average shift in the optimal search strategy depends on the pore characteristic size, while it is largely independent of the density of pores and the local configuration of the explored cluster (Fig. S2).

\subsection*{Scaling arguments}

\begin{figure}[!htbp]
\centering
\includegraphics[width=16cm]{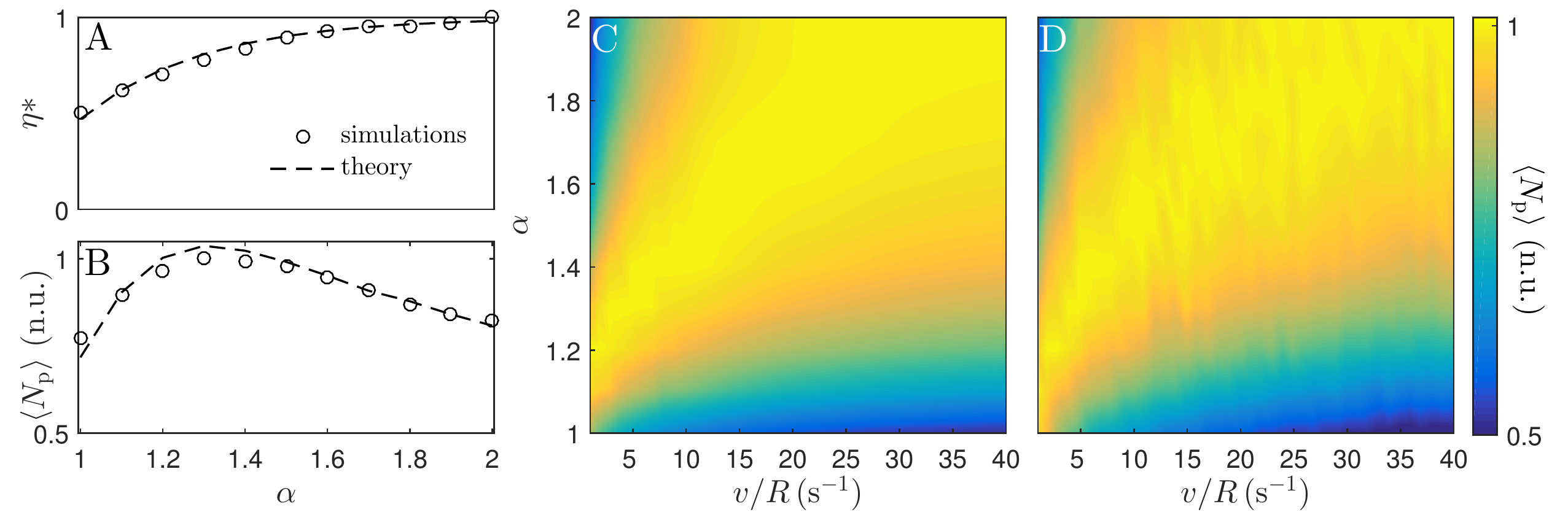}
\caption{\textbf{Influence of the topography on the optimal search strategy: comparison between simulated data and model}. (A) Normalized efficiency $\eta^*$: simulations (circles) and fit to Eq.~\ref{eta_norm} (dashed line). (B) Average number of caught targets in the porous environment as a function of $\alpha$: simulations (circles) and fit to the model (dashed line). (C) Model prediction and (D) simulated data of the average number of caught targets in the porous environment $\langle N_{\rm p}\rangle$ in normalized units (n.u.) as a function of $\alpha$ and normalized speed $v/R$ ($\gamma = 0.47$). }
\label{fig3}
\end{figure}

In order to formalize the shift in the optimal search strategy due to the topography of the environment, we define the efficiency $\eta$ of catching targets in the porous medium compared to the homogeneous case as
\begin{equation}\label{e5}
\eta(\alpha,v)
= {\langle N_{\rm p}(\alpha,v)\rangle \over \langle N_{\rm h}(\alpha,v)\rangle}.
\end{equation}
Since the mean square displacement of the active particle is of order $t^{3-\alpha}$ in a homogenous topography, self-intersections constitute a negligible fraction of the overall path for $\alpha < 2$, which is closely related to the fact that the Hausdorff dimension of a L\'evy process in the plane is equal to its exponent $\alpha$ \cite{blumenthal1960dimension}; this is also the case in the porous topography for run lengths just below the spatial cutoff (Fig.~\ref{fig2}C), which contribute with higher probability to the capture of new targets. As a consequence, to a first approximation, we obtain that the target capture rate is proportional to the average step length for a given topography and a given $\alpha$, so that
\begin{equation}\label{e6}
\eta(\alpha,v)
= {\langle \ell_{\rm p}(\alpha,v)\rangle \over \langle \ell_{\rm h}(\alpha,v)\rangle}
= \beta(v) \left[ 1 - (1-\gamma) \left( {T_{\rm c} \over t_{\rm c}} \right)^{-\alpha+1} \right],
\end{equation}
where $\beta(v) \in [0,1]$ is a function of the particle speed, $\gamma \in [0,1]$ is a constant, $T_{\rm c} = R_{\rm c}/v$, $t_{\rm c} = r_{\rm c}/v$, and $\langle \ell_{\rm p}(\alpha,v) \rangle$ and $\langle \ell_{\rm h}(\alpha,v) \rangle$ are the average step lengths in the porous and homogenous topography respectively (see Methods for their calculation).

While Eq.~\ref{e6} explicitly depends on the particle's speed $v$ through $\beta(v)$, the normalized efficiency $\eta^*$ defined as
\begin{equation}\label{eta_norm}
\eta^*(\alpha) 
= {\eta(\alpha,v) \over {\rm max}(\eta|v)} 
\approx {\eta(\alpha,v) \over \beta(v)}
=1 - (1-\gamma) \left( {T_{\rm c} \over t_{\rm c}} \right)^{-\alpha+1}
\end{equation}
is a universal curve that does not directly depend on $v$. Interestingly, from this equation, the geometrical meaning of $\gamma$ is apparent as the percentage of time that the particle spends running instead of being stuck at a boundary above the spatial cutoff $R_{\rm c}$.

Using Eq.~\ref{eta_norm}, we can therefore estimate the shift in the optimal search strategy due to the topography of the environment by finding the maximum of $\eta^*(\alpha) \langle N_{\rm h}(\alpha)\rangle$, i.e. only based on geometrical parameters ($t_{\rm c}$, $T_{\rm c}$ and $\gamma$) and the knowledge of the particle's behavior in a homogenous topography $\langle N_{\rm h}(\alpha)\rangle$. As it can be seen in Fig.~\ref{fig3}A, Eq.~\ref{eta_norm} reproduces very well the simulated data, being $\gamma$ the only fitting parameter in our case, and allows us to predict correctly the optimal value for the capture rate in the porous medium from $\langle N_{\rm h}(\alpha)\rangle$ (Fig.~\ref{fig3}B). By comparing model predictions (Fig.~\ref{fig3}C) with simulated data (Fig.~\ref{fig3}D), Figs.~\ref{fig3}C-D show how, once $\gamma$ is known, this simple model based on scaling arguments predicts correctly the optimal strategy in the porous medium at any particle speed. These figures show that, for a given $\gamma$, the speed at which the particle moves within the environment has also an effect on the optimal search strategy: for low values of speed, the optimal search strategy shifts towards the more ballistic case ($\alpha = 1$), as the particle tends to interact with the boundaries only at very long times, thus mostly moving as in an effectively homogenous environment; however, when $v$ increases, the optimal search strategy shifts more and more towards the Brownian case ($\alpha = 2$), since this case is the one that minimizes the interaction with the boundaries over time.

\begin{figure}[!htbp]
\centering
\includegraphics[width=12cm]{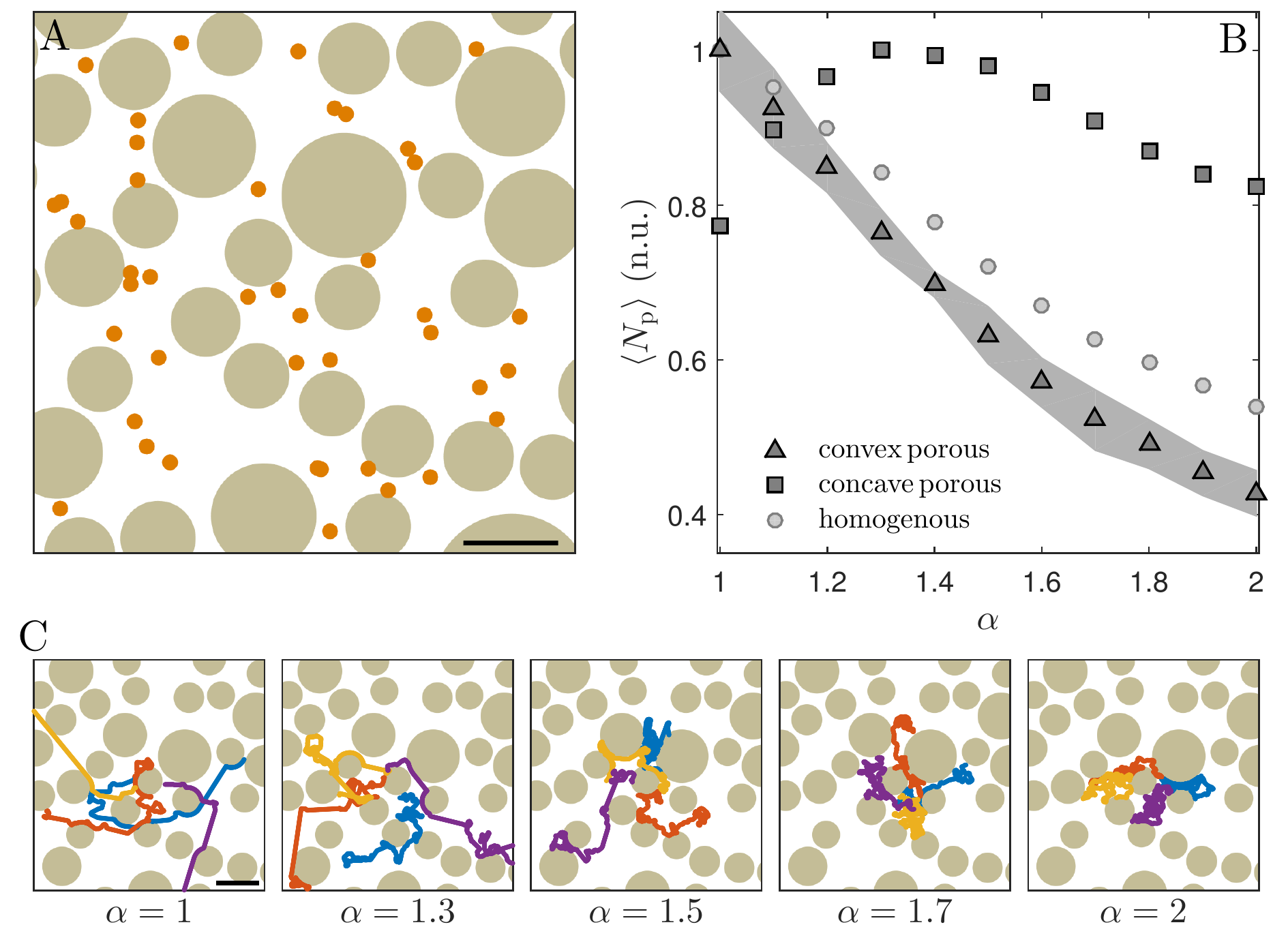}
\caption{\textbf{Convex vs. concave porous topography}. (A) A sample area of an extended two-dimensional convex porous medium where an active particle searches for uniformly distributed targets (dots). The porous medium is constituted by the space surrounding circular convex obstacles of average radius $R_{\rm p}$ ($R_{\rm p}/r_{\rm c} \approx 12.5$). (B) Average number of caught targets $\langle N_{\rm p} \rangle$ (triangles) as a function of $\alpha$ in normalized units (n.u.). The values are averaged over $1000$ $1$-hour trajectories and normalized to the maximum value at $\alpha = 1$. The grey shaded area represents one standard deviation around the average values. To directly compare with the homogenous and concave porous cases, the trends of Fig.~\ref{fig1}D and Fig.~\ref{fig2}B are also shown as circles and squares respectively. (C) Four different $1000$-s trajectories with a common origin are shown for different values of $\alpha$. All black scale bars corresponds to $50 \, R$.}
\label{fig4}
\end{figure}

For further confirmation of the fact that the shift in optimal search strategy is due to the upper spatial cutoff introduced by the topography of the environment, we now consider a porous medium with a convex topography instead of the concave topography considered previously (Fig.~\ref{fig2}A). In this topography, the particle searches for uniformly distributed targets within an interconnected space containing convex obstacles where there is no upper cutoff, i.e. $R_{\rm c} \to \infty$ (Fig.~\ref{fig4}A). Also in this case the average radius of the obstacles $R_{\rm p}$ is chosen so that $R_{\rm p}/r_{\rm c} \approx 12.5$. As expected, the optimal search strategy remains at $\alpha = 1$ (Fig.~\ref{fig4}B), as for a particle searching in a homogenous space (Fig.~\ref{fig1}D). In qualitative terms, these results can be interpreted by observing sample trajectories for various values of $\alpha$ (Fig.~\ref{fig4}C): in fact, as it can be appreciated from these trajectories, the convex porosity does not prevent the particles from moving ballistically over long distances when the value of $\alpha$ is decreased.

\subsection*{Search in the presence of Brownian diffusion}

\begin{figure}[!htbp]
\centering
\includegraphics[width=12cm]{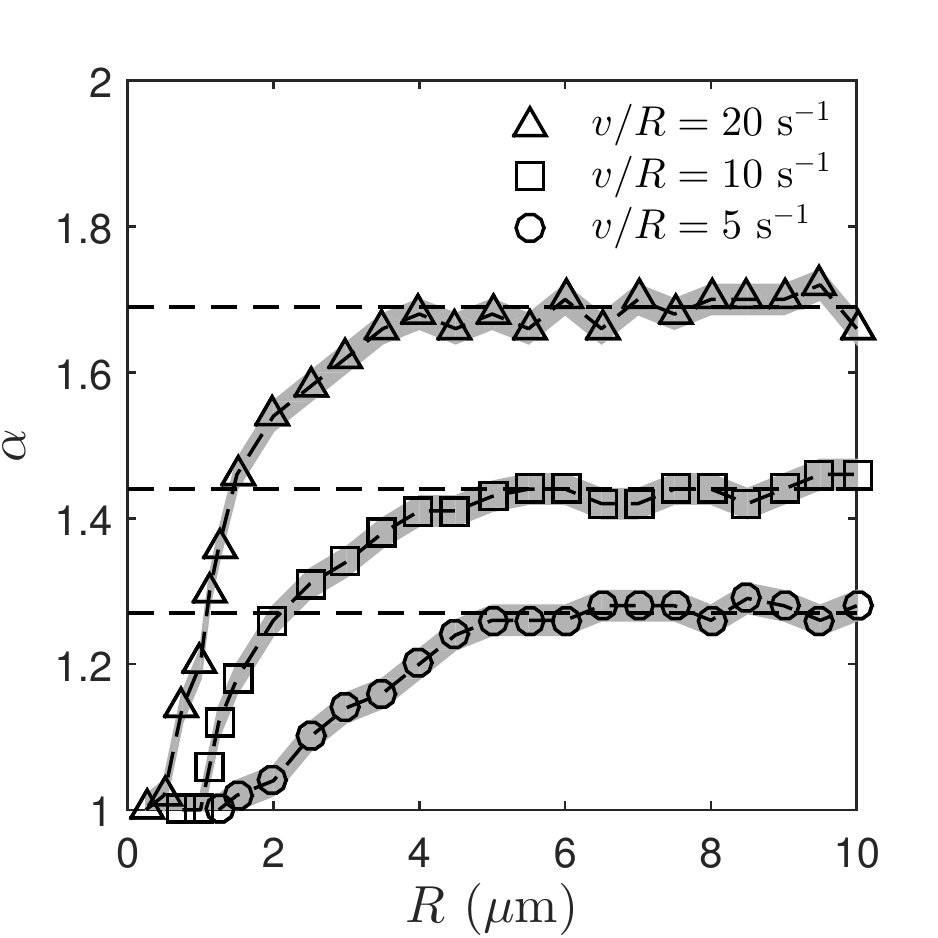}
\caption{\textbf{Optimal search strategy in the presence of Brownian diffusion}. (A) Optimal value of $\alpha$ as a function of the particle's radius $R$ for different values of $v$: $v = 5 R \, {\rm s}^{-1}$ (circles),  $v = 10 R \,{\rm s}^{-1}$ (squares),  $v = 20 R \,{\rm s}^{-1}$ (triangles). The dashed horizontal lines represent the optimal values in the absence of Brownian noise (Fig.~\ref{fig3}). (B-E) Shift of the optimal search strategy as a function of $\alpha$ in normalized units (n.u.) at the sample speed $v = 5 R \, {\rm s}^{-1}$ with decreasing values of the particle's radius $R$: (B) $ R = 5 \, {\rm \mu m}$, (C) $ R = 1 \, {\rm \mu m}$, (D) $ R = 0.5 \, {\rm \mu m}$ and (E) $ R = 0.1 \, {\rm \mu m}$. All values are averaged over 10000 1-hour trajectories. The grey shaded area represents one standard deviation around the average values.}
\label{fig5}
\end{figure}

The results shown so far apply to most length scales as long as properly rescaled to the particle's radius $R$. However, when $R$ approaches the micro- and nanoscale, Brownian diffusion starts playing a significant role in the translational and rotational motion of an active particle \cite{BechingerRMP2016,VolpeAJP2014}. In particular, while the translational diffusion of a particle scales with its inverse linear dimension ($\propto R^{-1}$), its rotational diffusion scales with its inverse volume ($\propto R^{-3}$). As a consequence of this volumetric scaling, as $R$ decreases, Brownian rotation randomizes any persistence in the particle's orientation due to the L\'evy strategy. Brownian diffusion becomes then an important parameter to consider when determining the optimal search strategy in a non-trivial topography for microscopic active particles, such as biological and artificial microswimmers (e.g. motile bacteria \cite{ArielNatCom2015,KorobkovaNature2004,TuPRL2005} and manmade micro- and nanorobots \cite{BechingerPRE2016}) moving in complex and disordered environments \cite{ChepizhkoPRL2013,ReichhardtPRE2014,PinceNatCom2016,BechingerRMP2016}. Fig.~\ref{fig5}A shows how the optimal search strategy (i.e. the optimal value of $\alpha$) varies as a function of the particle's radius, i.e. of the strength of the particle's translational and rotational Brownian diffusion coefficients. We focus again on the environment of Fig.~\ref{fig2}A, as it shows a clear deviation from the homogenous case (Fig.~\ref{fig1}). For a given $v/R$ (e.g. for $v/R = 5 \, {\rm s}^{-1}$), when $R$ is above a certain threshold value (e.g. $R \geq 5 \, {\rm \mu m}$ for $v/R = 5 \, {\rm s}^{-1}$, corresponding to a sufficiently weak rotational diffusion), the optimal strategy is the same as the one predicted in Fig.~\ref{fig3}C (Fig.~\ref{fig5}A and \ref{fig5}B). However, when $R$ decreases (entailing a stronger rotational diffusion), the optimal search strategy shifts towards $\alpha = 1$ (Fig.~\ref{fig5}A and \ref{fig5}C). This shift happens because the increased rotational diffusion leads to a reduction of the time that the particle spends at the boundaries. This effectively reduces the penalization that boundaries have on more ballistic strategies, thus allowing for the exploration of a greater inner area of the porous structure compared to more diffusive strategies. Reducing $R$ further, the optimal search strategy remains at $\alpha = 1$ although the relative efficiency over other $\alpha$ values decreases (Fig.~\ref{fig5}D). Finally, for even smaller values of $R$ (e.g. $R \leq 0.1 \, {\rm \mu m}$ for $v/R = 5 \, {\rm s}^{-1}$), the search process becomes effectively insensitive to the value of $\alpha$ (Fig.~\ref{fig5}E), as the increase in rotational diffusion makes persistent motion negligible \cite{BechingerRMP2016}.

\section*{Discussion}

Our results demonstrate the critical role played by the topography of the environment in determining the optimal search strategy for an active particle whose run lengths are drawn from a L\'evy distribution. In particular, the presence of physical boundaries, barriers and obstacles can introduce a cutoff on the distribution of steps that can penalize more ballistic strategies over more Brownian ones depending on different geometrical parameters connected to the topography of the environment and its interaction with the particle's motion. 

In our model, we assumed that the particle is performing a cruise search with continuous visibility for targets and perfect hitting probabilities. While we do not expect imperfect hitting probabilities to affect the optimality of the search strategy in our case as long as they affect all $\alpha$ values equally, other search scenarios might influence the optimal search strategy in a complex topography \cite{BenichouRMP2011}: for example, in the case of intermittent search strategies, where there is an alternation between phases of slow motion that allow the searcher to detect the targets and phases of fast motion during which targets cannot be detected; or in the case of a search strategy with in-built delays so that, after a target is caught, some time must elapse before the following target can be caught.

Another aspect that can influence the optimal search strategy is the interaction between the searcher and the obstacles encoded in the boundary conditions. In this work, we have implemented reflective boundary conditions, which implies that the searcher stays at the boundary until a random reorientation event makes it point away from the obstacle. This scenario is realistic at the macroscopic and microscopic scale, as for example both elementary robots and microswimmers (biological and non) have been reported to behave in this way \cite{DimidovANTS2016,BechingerRMP2016}. Alternatively, different responses can be considered in the presence of boundaries, when information obtained from sensing the surroundings for example leads to a voluntary switch in the strategy adopted by the searcher.

As the search time is generally a limiting factor in many realistic search scenarios \cite{BenichouRMP2011}, the searcher was allowed to explore the search space for a finite time in our simulations. Nevertheless, from a fundamental point of view, it would be interesting to study how the optimal search strategy is influenced by the topography of the environment in the limit of infinite search times. In the case of infinite searches, interesting behaviors could emerge as a result of the interplay between the fractal dimensionality of the searcher's trajectory and that of the environment in a porous topography at the percolation threshold or in a network of channels.

Our findings are mostly scale-invariant and only partially break down at the nanoscopic scale ($R \leq 1 \mu {\rm m}$) when rotational diffusion becomes predominant. One important implication of this is that different search strategies (i.e. different values of $\alpha$) will lead to similar outcomes for nanoscopic particles such as biomolecules and molecular motors moving in a two-dimensional space (Fig.~\ref{fig5}E). This issue can be overcome by reducing the dimensionality of the environment, for example by introducing a preferential direction of motion with molecular rails. In fact, L\'evy-type statistics emerge for molecular motors performing searches on polymer chains such as DNA \cite{LomholtPRL2005}, or on one-dimensional molecular rails such as microtubules \cite{ChenNatMat2015}. Similarly, increasing the dimensionality of the system to a three-dimensional space will alter the probability that the searcher goes back to the same point compared to a two-dimensional space, and thus its optimal search strategy in a complex three-dimensional environment can also be affected.

Our results are relevant for all random search problems where the searcher explores complex search spaces. Examples at various length scales include the rate of molecular encounters in the cytoplasm of cells, the localization of nutrients by motile bacteria in tissue or soil, the foraging of animals in patchy landscapes as well as search-and-rescue operations in ruins following natural disasters. Furthermore, similar dynamics could also be applied to optimize navigation in topologically complex networks \cite{RiascosPRE2012,GuoSciRep2016}.

\section*{Materials and Methods}

\subsection*{Numerical simulations}

In our numerical model, we consider active particles of radius $R$ performing a two-dimensional run-and-tumble motion according to the following equations: 
$$
\left\{\begin{array}{ll}
\displaystyle \frac{d}{dt} x(t) =  \displaystyle v \cos{\varphi_n} \\
\displaystyle \frac{d}{dt} y(t) =  \displaystyle v \sin{\varphi_n}\\
\end{array}\right.
$$\\
where $[x(t),y(t)]$ is the particle's position, $v$ is the particle's speed, and $ \varphi_n$ is the particle's orientation during the $n$-th time interval where $n=0,1,2,...$. The time intervals $t_n$ between changes of direction are drawn from a L\'evy distribution $P_{\alpha}(t)$ of exponent $\alpha \in [1, \, 2]$; only the absolute value of the number is considered. At the end of each time interval the particle orientation changes by a random angle according to a normal distribution with zero mean and standard deviation $\sigma_\varphi = \frac{\pi}{6}$. The initial position for the trajectory was randomly chosen within the medium according to a uniform distribution. The positions of the targets were randomized for each trajectory. Interactions with the walls were modeled using the boundaries conditions described in Ref.~\cite{VolpeAJP2014}. In the data in Fig.~\ref{fig5}, translational and rotational Brownian motion are included by adding three independent white noise processes ($W_x$ , $W_y$ and $W_{\varphi}$) to the equations of motion \cite{VolpeAJP2014}; in this set of simulations, the active particles are moving in an aqueous environment ($\eta = 0.001\, {\rm Nsm^{-2}} $, $T = 300\, {\rm K}$). 

\subsection*{Calculation of the average run length}

\subsubsection*{In a homogenous topography}

The average run length in a homogenous environment $\langle \ell_{\rm h}(\alpha)\rangle$ is 
$$
\langle \ell_{\rm h}(\alpha)\rangle
= \int_{0}^{\infty} \ell_{\rm h}(t) P_{\alpha}(t) dt = 
$$
$$
= \int_{0}^{t_{\rm c}(v)} \ell_{\rm h}(t) P_{\alpha}(t) dt + \int_{t_{\rm c}(v)}^{\infty} \ell_{\rm h}(t) P_{\alpha}(t) dt,
$$
where $t_{\rm c}(v) = r_{\rm c}/v$ represents the time that it takes for an active particle to travel a distance equal to its capture radius $r_{\rm c}$. Neglecting the first integral because it gives a small contribution to the average run length, we obtain
$$
\langle \ell_{\rm h}(\alpha)\rangle
\approx \int_{t_{\rm c}(v)}^{\infty} \ell_{\rm h}(t) P_{\alpha}(t) dt,
$$
which, using the asymptotic analytical form for $P_{\alpha}(t)$ in Eq.~\ref{e1}, can be calculated to be
\begin{equation*}
\langle \ell_{\rm h}(\alpha)\rangle
\approx v A(\alpha) \int_{t_{\rm c}(v)}^{\infty} t^{-\alpha} dt 
= v {A(\alpha) \over \alpha-1 }  t_{\rm c}^{-\alpha+1}.
\end{equation*}

\subsubsection*{In a porous topography}

The average run length in a porous environment $\langle \ell_{\rm p}(\alpha)\rangle$ is 
$$
\langle \ell_{\rm p}(\alpha) \rangle
= \int_{0}^{\infty} \ell_{\rm p}(t) P_{\alpha}(t) dt =  
$$
$$
= \int_{0}^{t_{\rm c}(v)} \ell_{\rm p}(t) P_{\alpha}(t) dt + \int_{t_{\rm c}(v)}^{T_{\rm c}(v)} \ell_{\rm p}(t) P_{\alpha}(t) dt + \int_{T_{\rm c}(v)}^{\infty} \ell_{\rm p}(t) P_{\alpha}(t) dt 
$$
where the integral has been divided into three parts delimited by the time cutoff at $t_{\rm c}$ and by that at $T_{\rm c} = R_{\rm c}/v$ calculated using the spatial cutoff introduced by the porous medium (Fig.~\ref{fig2}C). As for the homogenous case, the first integral gives a small contribution on the average run length as it is smaller than $r_{\rm c}$. As such it can be neglected, so that 
$$
\langle \ell_{\rm p}(\alpha)\rangle
\approx \int_{t_{\rm c}(v)}^{T_{\rm c}(v)} \ell_{\rm p}(t) P_{\alpha}(t) dt + \int_{T_{\rm c}(v)}^{\infty} \ell_{\rm p}(t) P_{\alpha}(t) dt.
$$
We can now treat these two integrals using the fact that, due to the interaction with the boundaries, $\langle \ell_{\rm p}(t) \rangle \leq \langle \ell_{\rm h}(t) \rangle$ at any given time $t$ taken from the distributions of Eq.~\ref{e1} (Fig.~\ref{fig1}B). In general, $\langle \ell_{\rm p}(t) \rangle = C(t,\alpha,v) \langle \ell_{\rm h}(t) \rangle$, where $C \in [0,1]$ is a multivariable function. To simplify the analysis, we introduce the following approximation: $C(t,\alpha,v) = \beta(v)$ for $t \in [t_{\rm c},T_{\rm c}]$ and $C(t,\alpha,v) = \gamma \beta(v)$ for $t \in [T_{\rm c}, \infty)$, where $\beta \in [0,1]$ is a speed-dependent constant and $\gamma \in [0,1]$ is a prefactor related to the topography of the environment. This approximation allow us to determine the decrease of the average run length in the porous environment over the homogenous case by estimating the decrease of the area of the integral before and after the time cutoff at $T_{\rm c}$ (Fig.~\ref{fig2}C), and thus to treat differently the distribution of the run lengths in the porous environment $P_{{\rm \rm p},\alpha}(\ell)$ (Eq.~\ref{e3}) in the two time intervals. We therefore obtain for the two integrals:
\begin{equation*}
\int_{t_{\rm c}(v)}^{T_{\rm c}(v)} \ell_{\rm p}(t) P_{\alpha}(t) dt \approx \beta(v) v {A(\alpha) \over \alpha-1} \left( t_{\rm c}^{-\alpha+1} - T_{\rm c}^{-\alpha+1} \right)
\end{equation*}
and
\begin{equation*}
\int_{T_{\rm c}(v)}^{\infty} \ell_{\rm p}(t) P_{\alpha}(t) dt \approx \gamma \beta(v) v {A(\alpha) \over \alpha-1} T_{\rm c}^{-\alpha+1}
\end{equation*}
Summing these two integrals we obtain:
\begin{equation*}
\begin{split}
\langle \ell_{\rm p}(\alpha) \rangle 
& \approx \beta(v) v {A(\alpha) \over \alpha-1}  \left( t_{\rm c}^{-\alpha+1} - T_{\rm c}^{-\alpha+1} \right) + \gamma \beta(v) v {A(\alpha) \over \alpha-1} T_{\rm c}^{-\alpha+1} = \\
& = \beta(v) v {A(\alpha) \over \alpha-1} t_{\rm c}^{-\alpha+1}
\left[ 1 - (1-\gamma) \left({T_{\rm c}\over t_{\rm c}}\right)^{-\alpha+1} \right]
\end{split}
\end{equation*}

\subsection*{Dataset}

Dataset available at https://doi.org/10.6084/m9.figshare.5488756.v1

\section*{Acknowledgements}

We acknowledge Jan Wehr for useful discussions on the mathematical part of the manuscript, and Er\c{c}a\u{g} Pin\c{c}e, Mite Mijalkov, Geet Raju, and Sylvain Gigan for useful discussions in the initial stages of the project. We also acknowledge the COST Action MP1305 "Flowing Matter" for providing several meeting occasions. Giorgio Volpe acknowledges funding from the Wellcome Trust [under Grant 204240/Z/16/Z]. Giovanni Volpe acknowledges funding from the European Research Council (ERC Starting Grant ComplexSwimmers, grant number 677511).

\setcounter{figure}{0}

\makeatletter 
\renewcommand{\thefigure}{S\@arabic\c@figure}
\makeatother

\begin{figure}[!htbp]
\centering
\includegraphics[width=12cm]{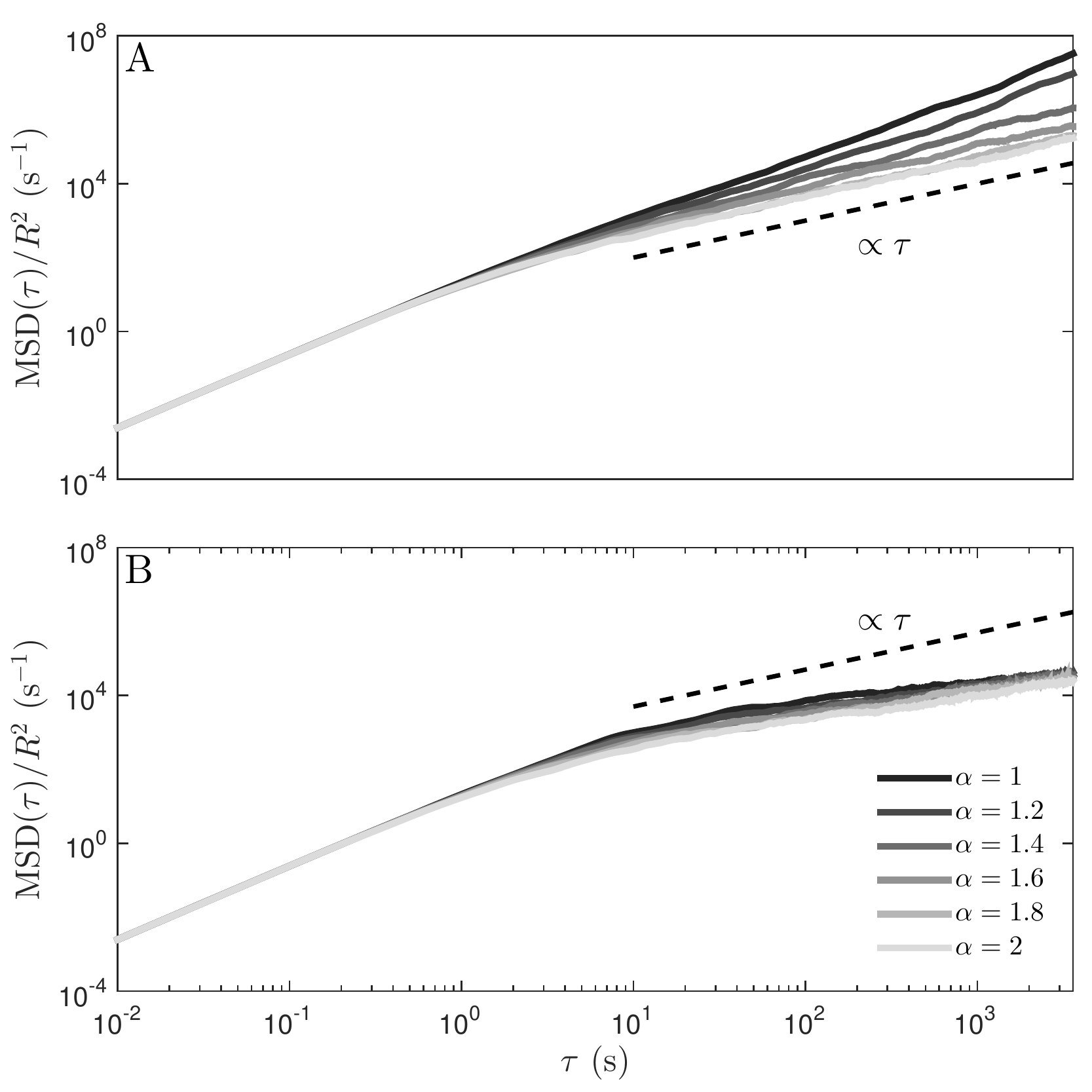}
\caption{\textbf{Searcher's MSDs in different topographies}. Average mean square displacements (MSDs) of searchers for different values of $\alpha$ (A) in a homogenous topography, showing superdiffusive behavior, and (B) in a porous topography, showing subdiffusive behavior. The dashed line represents diffusive behavior. Each MSD curve was obtained as an ensemble average over 100 1-hour trajectories.}
\label{figS1}
\end{figure}

\begin{figure}[!htbp]
\centering
\includegraphics[width=12cm]{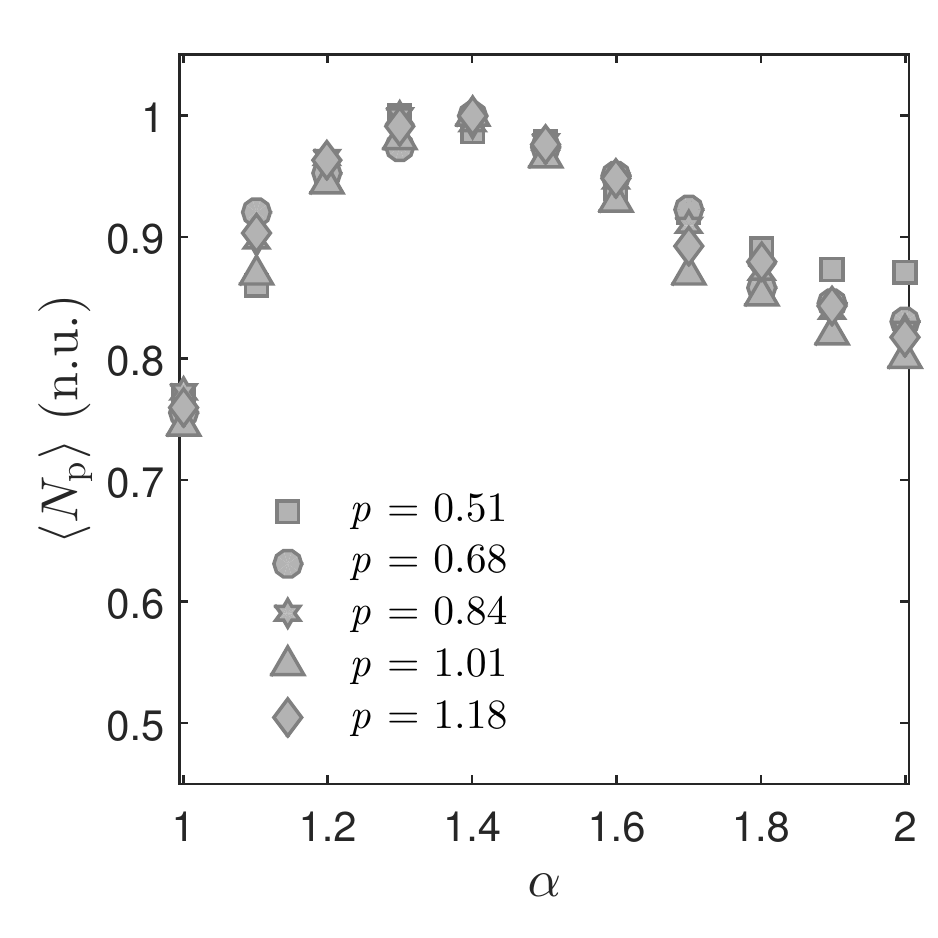}
\caption{\textbf{Optimal search strategy in a porous topography for different densities of pores}. Average number of caught targets as a function of $\alpha$ in normalized units (n.u.) for different pore densities $p$. All data collapse on the same curve. The case $p = 0.84$ corresponds to the curve in Fig.~\ref{fig2}B. At $p = 1$, the porous structure is at percolation and the area occupied by the pores is approximately $77\%$ of the total area. Every curve is averaged over 1000 1-hour trajectories and normalized to its maximum value.}
\label{figS2}
\end{figure}

\end{document}